# Exploring the Interaction of BeS Monolayer and Lung Disease Biomarkers: Potential Material for Biosensing Applications

Sudipta Saha and Md. Kawsar Alam[1]

Department of Electrical and Electronic Engineering, Bangladesh University of Engineering and Technology, Dhaka-1205, Bangladesh

## Abstract

Considerable attention has been directed towards the prognosis of lung diseases primarily due to their high prevalence. Despite advancements in detection technologies, current methods such as computed tomography, chest radiographs, bold proteomic patterns, nuclear magnetic resonance, and positron emission tomography still face limitations in detecting diseases related to the lungs. Consequently, there is a need for swift, non-invasive and economically feasible detection methods. Our study explores the interaction between BeS monolayer and breathe biomarkers related to lung disease utilizing the density functional theory (DFT) method. Through comprehensive DFT analysis, including electronic properties analysis, charge transfer evaluations, work function, optical properties assessment and recovery times, the feasibility and efficiency of BeS as a VOC (volatile organic compound) detection are investigated. Findings reveal significant changes in bandgap upon VOC adsorption, with notable alteration in work function for selective compounds. Optical property analyses demonstrate the potential for selective detection of biomarkers within specific wavelength ranges. Moreover, the study evaluates the impact of electric fields and strain on VOC-2D BeS interaction. Furthermore, the desorption of these VOCs from the BeS surface can be achieved through a heating process or under the illumination of UV light. This feature enables the reusability of the 2D material for biosensing applications. These findings highlight the potential of the BeS monolayer as a promising material for the sensitive and selective detection of breath biomarkers related to lung disease.

---

[1] Corresponding Author. Email: kawsaralam@eee.buet.ac.bd; kawsar.alam@alumni.ubc.ca

## Introduction

Lung diseases encompass many conditions that affect the lungs, including infections, chronic respiratory diseases, and lung cancer. These diseases can significantly impact an individual's respiratory function and overall health. Common examples of lung diseases include asthma, chronic obstructive pulmonary disease (COPD), pneumonia, and lung cancer. Notably, lung cancer has consistently held the title of the most frequently diagnosed cancer worldwide for several decades [1]. In recent years, technologies like computed tomography (CT) [2,3], chest radiographs (CXRs) [4], positron emission tomography (PET) [5], bold proteomic patterns [6], nuclear magnetic resonance (NMR) [7], magnetic resonance imaging (MRI) [5] and have gained extensive usage in clinical settings for the detection of various lung diseases. Despite their effectiveness, these methods come with significant drawbacks; for instance, conventional X-rays (CXR), CT scans, and PET scans entail radiation exposure risks [8,9], high operational costs [10,11] and limitations in providing a conclusive diagnosis only at advanced stages of the particular disease. Thus, these techniques still fall short in detecting the early stages of the disease and fail to enhance survival rates for patients notably [12]. Due to these limitations, a swift, non-invasive, and economically feasible means of detecting lung diseases holds significant value. Researchers are presently investigating novel methods to detect diseases in their early stages. Exhaled human breath emerges as a promising, cost-effective, and swift approach for detection [13,14]. Volatile organic compounds (VOCs), including hydrocarbons like isoprene ($C_5H_8$), hydrocarbon derivatives such as 2-propenal ($C_3H_4O$), 4-hydroxyhexanal ($C_6H_{10}O_2$) and acetone ($C_3H_6O$), along with diverse aromatic hydrocarbons like benzene ($C_6H_6$), act as biomarkers signifying the extent of lung diseases in humans (**Figure 1**) [15-18]. Recent progress in utilizing 2D materials for detecting biomarkers has received considerable attention [19-21]. The application of 2D materials, characterized by their high surface-to-volume ratio, shows promising prospects for enabling sensing at concentrations as low as a few ppb (parts per billion), thus achieving remarkable sensitivity [22,23]. In prior research, it was demonstrated that $MoS_2$ [24], $MoSe_2$ [25], $SnS_2$ [26] and $WSe_2$ [27] were proposed as biosensors for detecting lung cancer through exhaled breath analysis. Additionally, various forms of metal doping have been incorporated into $MoS_2$, $MoSe_2$, and $SnS_2$ to enhance their sensing capabilities [24,26,28-30]. A first principle investigation into the functionalization of titanium carbide $Ti_3C_2$ MXenes was conducted to observe the detection of biomarkers related to lung cancer [31]. In VOC detection, group II-VI monolayers such as BeS, MgO and BeSe have received little attention recently. Theoretical predictions suggest that the BeS monolayer features a considerably wide bandgap and notable electronic and optical properties [32,33]. Furthermore, the BeS monolayer has demonstrated potential in gas sensing applications [34]. However, utilizing the BeS monolayer for identifying lung disease biomarkers is yet to be explored. The remarkable optical and electronic properties of the BeS monolayer, combined with its effectiveness in gas sensing and capture, have inspired the present study to investigate its prospect for sensing application of diseases related to lungs utilizing this 2D material. The interaction process has been thoroughly examined by assessing parameters including adsorption energy, bond or interaction distance, electron density difference and charge transfer. Furthermore, analysis of the PDOS (partial density of states) and electronic band structure has been done to study the electronic properties of the VOC adsorbed system. Exploring the feasibility of using this material for biosensing application, various parameters such as conductivity, work function, reflectivity, absorbance and recovery time for all VOC adsorbed systems have been computed. Additionally, interaction has been observed under the influence of electric field and strain. The following sections provide detailed explanations of the computational methodology, procedures, and study results.

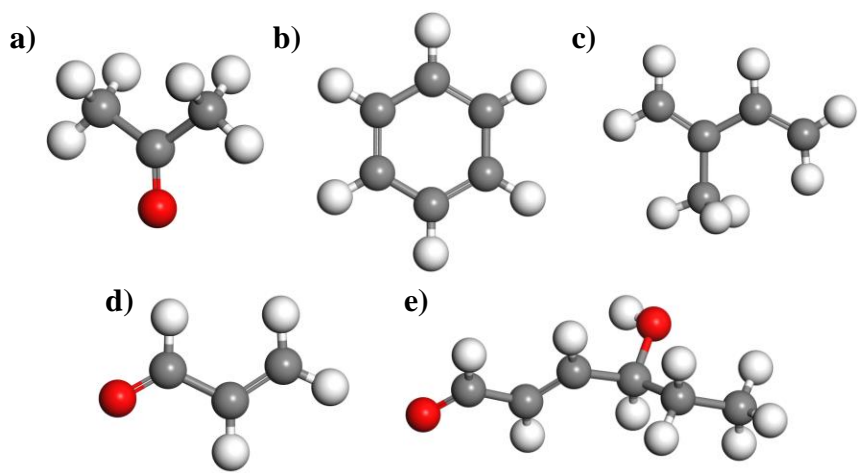

**Figure 1.** Volatile organic compounds linked to lung diseases (a) acetone, (b) benzene, (c) isoprene, (d) 2-propenal and (e) 4-hydroxy hexanal. Here, white, grey, and red spheres represent hydrogen, carbon, and oxygen atoms, respectively.

**Computational Description and Methodology**

In this study, density functional theory (DFT) calculations were performed utilizing the DMol$^3$ [35] and CASTEP [36] modules of Materials Studio. Various computational analyses were conducted, including geometry optimization, electronic band structure analysis, density of states calculation, and Hirshfeld charge analysis using the DMol$^3$ module. In contrast, phonon dispersion, electron density difference, and optical property assessment were performed using the CASTEP module. The study was conducted with a 3×3 supercell extracted from bulk BeS crystal, with a 21Å vacuum slab to mitigate undesired interactions due to the periodic images. In sensing applications, the change in the band gap is more significant than its absolute value. So, GGA-PBE functional [37] and DFT-D correction methods proposed by Grimme [38] were selected due to their computational efficiency and consistency with previous DFT studies focusing on gas and bio-sensing [24,39]. The geometry of the BeS monolayer was optimized using the DMol$^3$ module until force and energy met the convergence thresholds of $1.0 \times 10^{-4}$ Ha per Å and $1.0 \times 10^{-7}$ Ha between two consecutive steps. The Fermi level smearing parameter of 0.005 Ha was utilized alongside a Monkhorst Pack grid with dimensions of 16×16×1. A global orbital cutoff of 5.0 Å was employed, complemented by the numerical atomic orbital-based Double Numerical Polarized (DNP) basis set. Following this setup, VOCs were introduced to primary adsorption sites to ascertain the most stable and optimal adsorption site. Each volatile organic compound was investigated across four distinct adsorption sites, which encompassed the region between the Be-S bond (designated as the bridge site), the top of the Be atom, the top of the S atom, and the midpoint of the hexagon (referred to as the hollow site). Following the adsorption process, the structures underwent relaxation utilizing the same parameters employed in the structural relaxation of the pristine monolayer. The adsorption energy of VOC-adsorbed systems and charge transfer between the VOC and the BeS was calculated using the following equations [40]:

$$E_{ad} = E(adsorbate + BeS) - E(BeS) - E(adsorbate) \tag{1}$$

$$\Delta\rho = \rho(adsorbate + BeS) - \rho(BeS) - \rho(adsorbate) \tag{2}$$

Here, $E_{ad}$ is the adsorption energy, $E(adsorbate)$ is the energy of the VOC, $E(adsorbate + BeS)$ is the energy of the VOC-adsorbed BeS monolayer and $E(BeS)$ corresponds to the energy of the BeS monolayer. A more negative value of $E_{ad}$ suggests a robust exothermic interaction and stable adsorption under thermal conditions. In the subsequent analysis, only the configurations with the highest energy favorability were considered.

$\rho(adsorbate + BeS), \rho(BeS), \rho(adsorbate)$ denotes the charge on the VOC-adsorbed BeS monolayer, pure BeS monolayer and the VOC, respectively

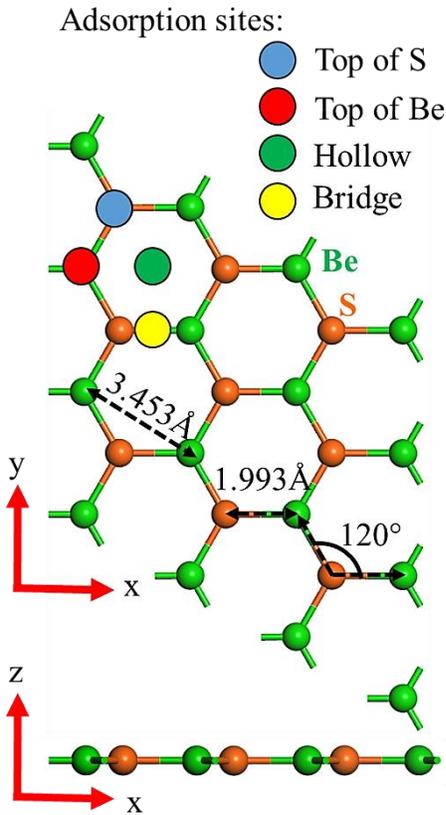

**Figure 2.** Optimized structure of pristine BeS monolayer's top view and side view. Four adsorption sites on the pristine BeS monolayer are indicated by blue (Top of S), red (Top of Be), green (Center of the hexagon) and red (Between Be-S bond) circles.

# Result and discussion

## A. Benchmarking Simulation Methodology and Parameters

At first, electronic and structural characteristics of pristine BeS monolayer were computed to establish a benchmark for the simulation methodology. **Figure 2** displays the top views and side views of the pristine monolayer. BeS monolayer exhibits a planar arrangement with a lattice constant of 3.459Å, a bond length of 1.993Å and a bond angle of 120°. The band structure and DOS of the pristine monolayer are shown in **Figure 3 (a & b)**. It was found that the BeS monolayer has a band gap of 4.529 eV. These findings align with previous theoretical investigations [34,41,42]. As there is no negative frequencies present in the **Figure 3(c)** the pristine monolayers of BeS can be considered stable.

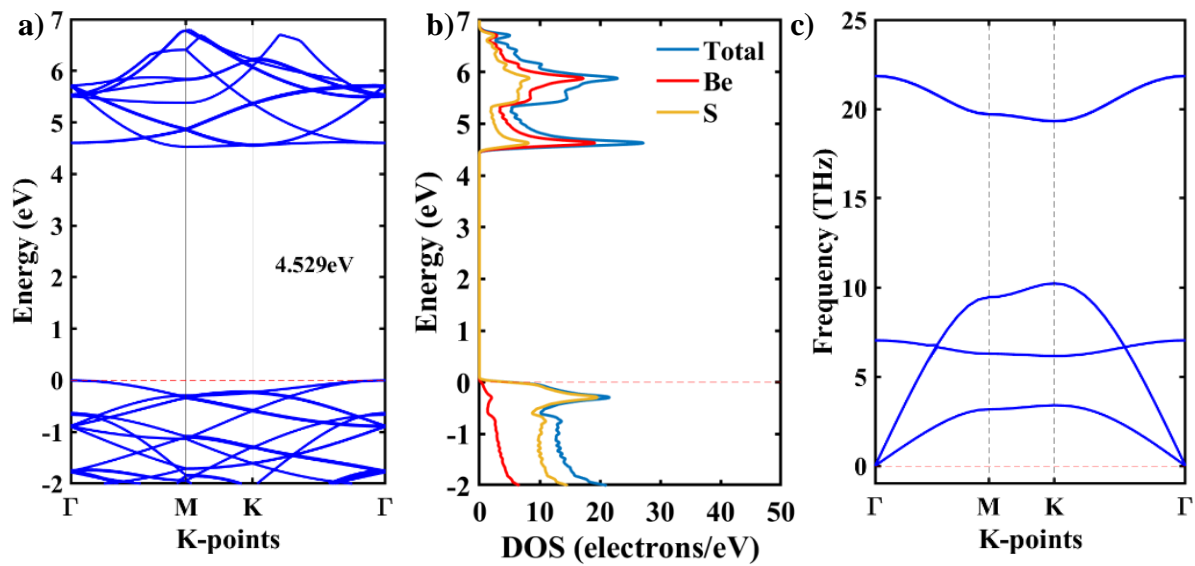

**Figure 3.** BeS monolayer's (a) band structure (path Γ-M-K-Γ), (c) density of states and (d) phonon dispersion of pristine BeS monolayer

## B. VOCs Adsorption on BeS Monolayer

As mentioned earlier, four individual adsorption locations for each VOC, including the center of the hexagon (referred to as the hollow site), the area between the Be-S bond (referred to as the bridge site), the top of the Be atom and the top of the S atom was considered. Only the most stable and energy-favorable adsorption geometries of all BeS analyte systems are demonstrated in **Figure 4.** Different VOCs display affinities for different adsorption sites.

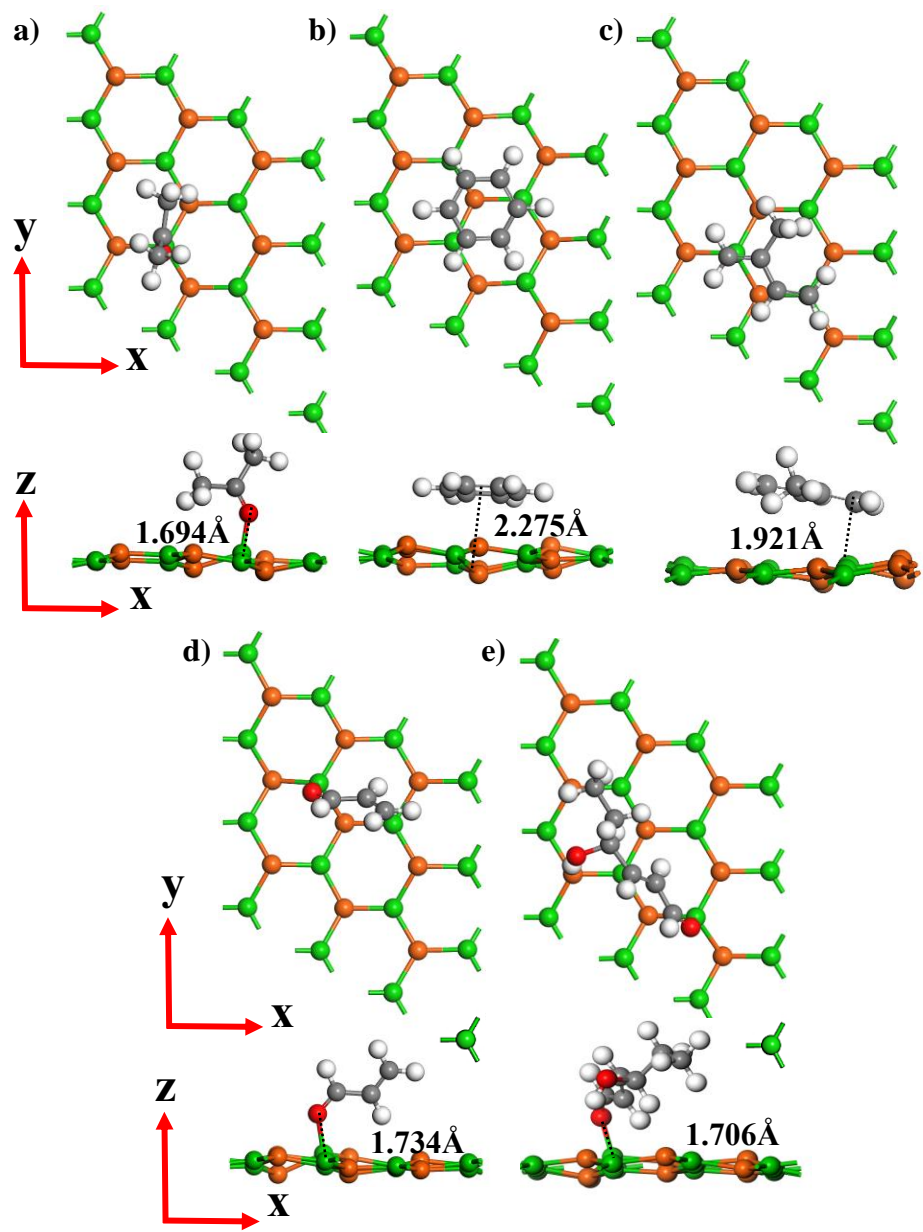

**Figure 4.** The most optimized structure (top view and side view) for (a) acetone, (b) benzene, (c) isoprene, (d) 2-propenal and (e) 4-hydroxy hexanal (4HHE) on pristine BeS surface. Green, orange, grey, white, and red spheres represent Be, S, C, H, and O atoms, respectively.

Observations of fluctuations in the shortest observed distance (D) between the VOC molecules and the pristine monolayer surface have been noted. The bonding process between the monolayer and VOC molecules is influenced by intermolecular distance, which plays a significant role in defining the adsorption process. In **Table 1,** adsorption energy ($E_{ad}$), adsorption distance (D), charge transfer (Q), band gap ($E_g$), band gap change compared to pristine (%$\Delta E_g$), work function ($\Phi$), recovery time and desorption temperature ($T_D$) have been summarized for all BeS analyte systems. If the value of Hirshfeld charge transfer Q(e) is positive (negative), then VOC is considered to be a donor (acceptor). All the charge density difference plots are provided in the supplementary document **Figure S1.** Recovery time and desorption temperature are indicators of the reusability of the biosensor. The adsorption of any VOCs may occur in two ways: physisorption and chemisorption. The $E_{ad}$ is reduced in physisorption, binding the VOC molecule to the substrate through weak vdW (van der Waals) forces, with minimal charge transfer between the substrate monolayer and VOC molecule. Conversely, chemisorption involves a chemical interaction between the VOC molecule and substrate material, characterized by high adsorption energy and substantial charge transfer, leading to a strong charge transfer between the substrate monolayer and VOCs. There is a clear correlation between adsorption energy and material sensitivity. A higher adsorption energy indicates a stronger interaction between the monolayer and VOC molecules, enhancing the material's sensitivity. On the contrary, strong interaction impedes the desorption of VOC molecules from the material's surface, presenting obstacles to the reusability of gas sensors [43,44].

Regarding the adsorption of acetone on BeS, the preferred adsorption site is deemed to be atop the Be atom, as depicted in **Figure 4(a).** A bond was formed between the O atom of VOC and the Be atom of the pristine monolayer, and the bond length was observed to be 1.694Å. Table 1 shows that the charge transfer between the VOC and pristine is 0.343|e|, and the adsorption energy is -0.845eV. The charge transfer can be visualized from **Figure S1(a)**, where a negative charge was accumulated near the O atom, and the charge was depleted from the surface of pristine BeS. These findings are indications of strong chemisorption. From **Figure 4(b),** it can be observed that for the adsorption of benzene on BeS, the most stable site was found to be on top of the S atom. The distance between the S atom and C atom was 2.275Å, and the charge transfer was 0.055|e|. Adsorption energy was -0.528eV. Lower adsorption energy, smaller charge transfer, and higher adsorption distance indicate that the interaction was of a physisorption nature. For isoprene in **Figure 4(c)**, the adsorption distance was 1.921Å on top Be atom, charge transfer was 0.258|e|, and adsorption energy was -0.551eV and it was physisorbed. For the adsorption of 2-propenal as shown in **Figure 4(d)**, the adsorption distance was 1.734Å, charge transfer was 0.282|e|, and adsorption energy was -0.763 eV. In this scenarios, a bond was established, and the interaction between the VOCs and pristine BeS was of a chemisorption nature. Highest adsorption energy of -1.141eV was observed in case of adsorption of 4-hydroxy hexanal (4HHE) on BeS monolayer and bond length was observed to be 1.706Å. From **Table 1,** it can be also observed that charge transfer between 4HHE and BeS monolayer is also significant.

**Table 1.** Adsorption energy ($E_{ad}$), Hirshfeld charge transfer (Q), adsorption distance (D) observed bandgaps ($E_g$) for all adsorbed systems and corresponding bandgap change with respect to the pristine BeS, work function ($\Phi$), recovery time of the VOC-BeS system and desorption temperature ($T_D$)

| System | $E_{ad}$ (eV) | Q (e) | D (Å) | $E_g$ (eV) | %$\Delta E_g$ | $\Phi$ (eV) | Recovery Time @298K (s) | Recovery Time @298K and UV(s) | $T_D$ (K) |
|---|---|---|---|---|---|---|---|---|---|
| Pristine BeS | - | - | - | 4.529 | - | 6.56 | - | - | - |
| $C_3H_6O$ - BeS | -0.845 | 0.343 | 1.694 | 2.581 | 43.01 | 5.93 | 19.72 | 0.019 | 411.16 |
| $C_6H_6$ - BeS | -0.528 | 0.055 | 2.275 | 4.292 | 5.23 | 6.59 | $8.58 \times 10^{-5}$ | $8.58 \times 10^{-8}$ | 294.45 |
| $C_5H_8$ - BeS | -0.551 | 0.258 | 1.921 | 3.534 | 21.97 | 6.28 | $2.08 \times 10^{-4}$ | $2.08 \times 10^{-7}$ | 315.83 |
| $C_3H_4O$ - BeS | -0.763 | 0.482 | 1.734 | 1.512 | 66.62 | 6.01 | 0.81 | $8.1 \times 10^{-4}$ | 309.83 |
| $C_6H_{10}O_6$ - BeS | -1.141 | 0.438 | 1.706 | 1.920 | 57.61 | 6.23 | $1.97 \times 10^6$ | $1.97 \times 10^3$ | 334.38 |

From **Table 2**, it can be observed that charge transfer between the VOCs and pristine BeS was significantly higher for acetone, isoprene, and 2-propenal compared to that of other monolayers for sensing. Charge transfer for benzene was also higher compared to most of the $Ti_3C_2$ MXenes.

**C. The stability analysis**

Before investigating the electronic and optical properties of different VOC-adsorbed systems, the phonon spectra of these systems were analyzed to ensure the system's stability. As illustrated in **Figure 5**, no negative frequencies were observed in the phonon spectra of any VOC molecules. Also, thermodynamic stability of each VOC adsorbed system can be verified from **Figure S2.**

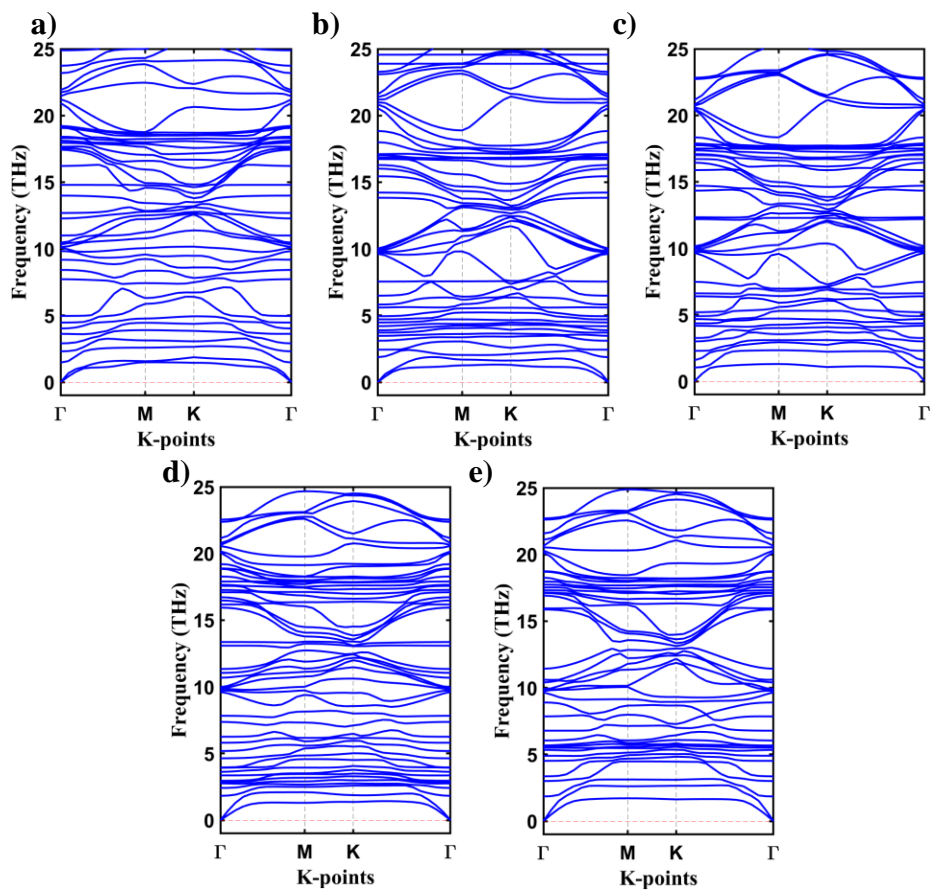

**Figure 5.** Phonon dispersion of (a) acetone , (b) benzene , (c) isoprene, (d) 2-propenal and (e) 4-hydroxy hexanal (4HHE) adsorbed structure.

### D. The Electronic Properties of BeS-analyte System

The analysis involved studying the band structure and partial density of states (PDOS) of both the pristine monolayer and the BeS monolayers adsorbed with various VOCs, as the adsorption of different VOCs induces modifications in the electronic properties of the monolayer. **Figure 6(a)** illustrates the band structure of acetone adsorbed BeS. The bandgap was observed to be 2.581eV. It is significantly reduced compared to the band gap of pristine BeS. The PDOS plot in **Figure 6(a)** helps better understand the role played by the adsorbate molecules in altering the electronic structure of the BeS monolayer. Strong hybridization was contributed by the p orbital of C and O atoms, and the s orbital of the H atom was observed near the 2.5eV region. Hybridization occurs within the forbidden band region of the pristine BeS monolayer, resulting in a substantial 43.01% reduction in the bandgap.

Similar phenomena are also observed when isoprene, 2-propenal and 4HHE is adsorbed on BeS monolayer. From **Figure 6(c-e),** it can be observed that the change in that bandgap is 21.97%, 66.62% and 57.61% due to the adsorption of isoprene, 2-propenal and 4HHE respectively. It is worth mentioning that the bandgap of the BeS-isoprene system is indirect.

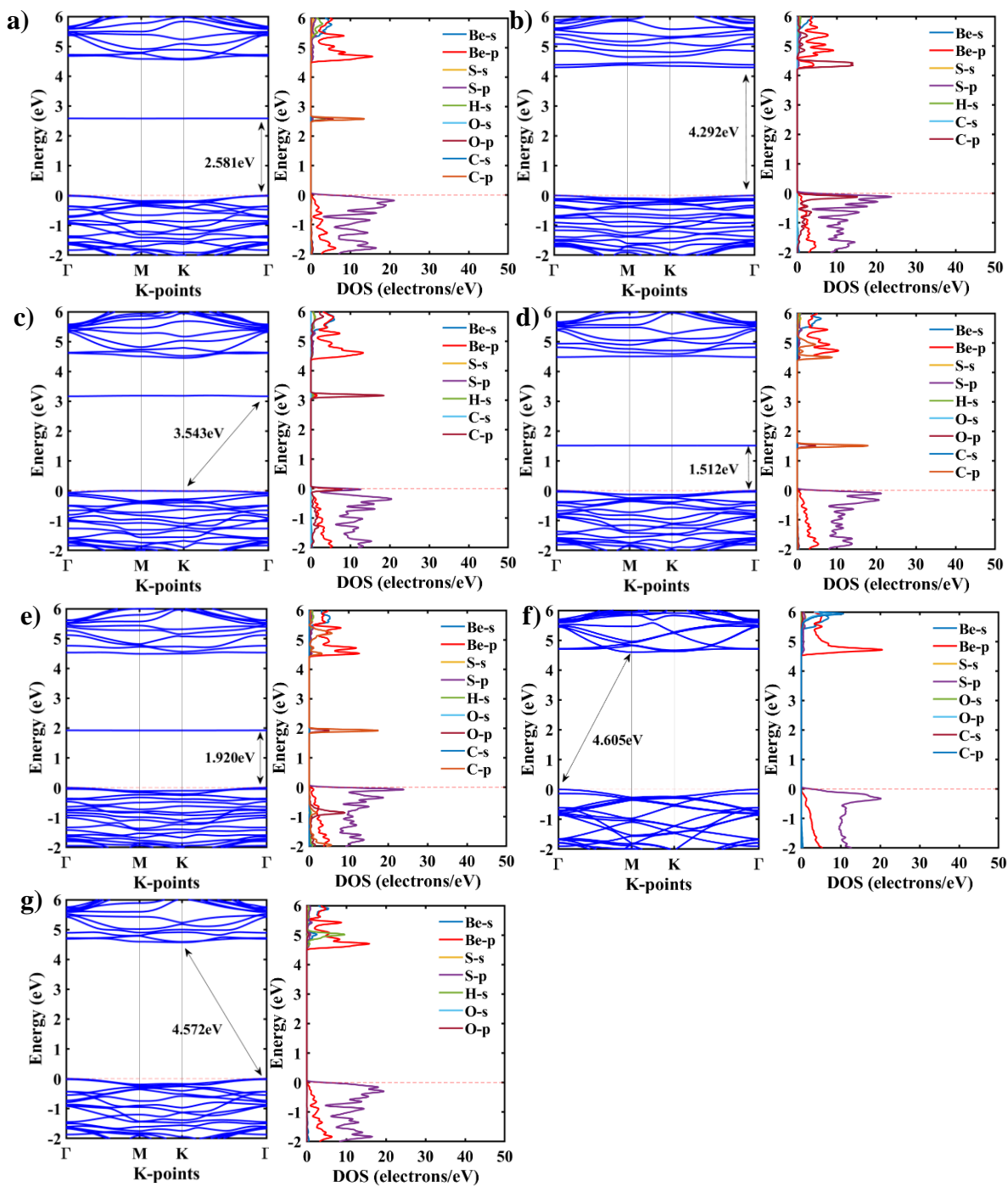

**Figure 6.** Band structure (path Γ-M-K-Γ) and density of states for (a) acetone, (b) benzene, (c) isoprene, (d) 2-propenal, (e) 4-HHE, (f) $CO_2$ and (g) $H_2O$.

By looking at **Figure 6(c-e)**, hybridization between the p orbital of C and O atoms causes defects at 3eV, 1.5eV and 2eV region, respectively. The presence of defect states leads to a reduction in the bandgap of the VOC adsorbed system in comparison to the pristine BeS monolayer.

However, **Figure 6(b)** depicts that when benzene was adsorbed on the pristine BeS, the change in electronic structure was minimal. Analyzing the PDOS plot shown in **Figure 6(b),** the contribution of the p orbital of the C atom near the conduction band minima and valence band maxima reduced the bandgap by 5.23%.

### E. The Optical Properties of BeS-analyte System

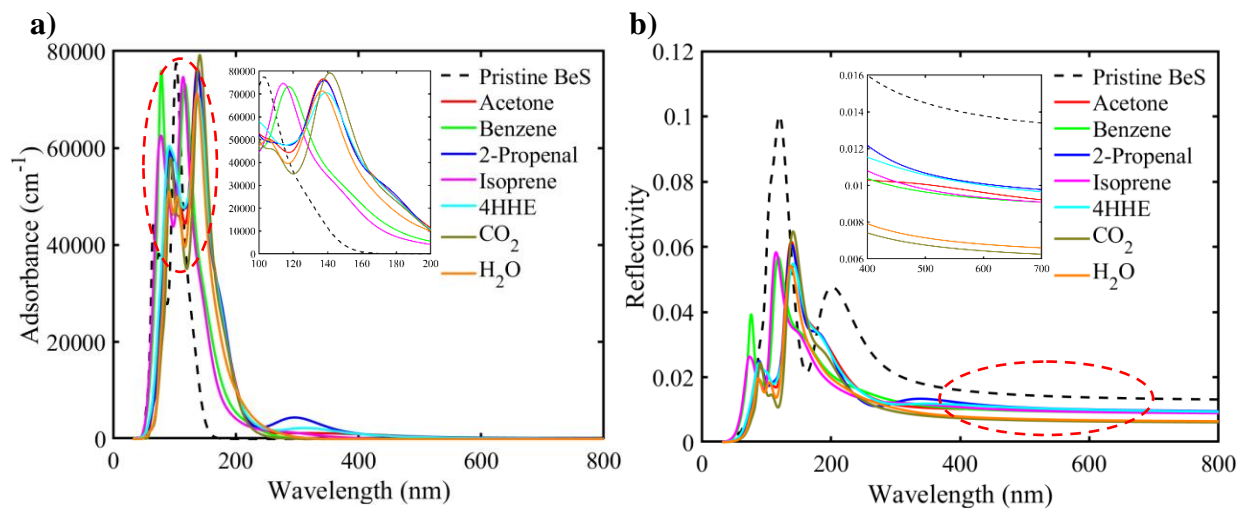

**Figure 7.** Calculated (a) Absorbance and (b) reflectivity for VOCs and interfering molecules adsorbed on BeS monolayer

In recent years, optical biosensors have gained attention due to their ability to detect changes in a material's optical properties caused by adsorbate molecules [45]. This phenomenon holds promise for selectively sensing volatile organic compounds (VOCs), offering advantages over traditional electrical sensors like immunity to electromagnetic interference and high sensitivity [46]. Additionally, optical biosensors can detect a wide range of molecules by tuning the sensor's frequency. Consequently, exploring whether the BeS monolayer could serve as an effective optical biosensing material with tailored optical properties for different VOCs is crucial. To assess this, absorption coefficient and reflectivity calculations have been performed for various VOC and interfering molecules, such as $H_2O$ and $CO_2$ adsorbed systems, aiming to determine the BeS monolayer's suitability for selective VOC sensing applications. When light is incident on the VOC-adsorbed BeS structure, major absorption peaks occur between 100~200nm region, as seen in **Figure 7(a)**. The benzene and isoprene adsorbed structures exhibit peak absorbance around 120nm, enabling their selective detection. However, other volatile organic compound (VOC) molecules, as well as interfering molecules, demonstrate peak absorbance near 140nm, posing challenges for selectivity. Moreover, no light absorption occurs in the visible range.

Reflectivity is a critical optical property for consideration in two-dimensional biosensors due to their reduced thickness. Within the visible region (400nm to 700nm), all volatile organic compounds exhibit over 20% reflectivity, distinguishing them from interfering molecules as shown in **Figure 7(b)**.

### E. Work Functions of BeS-Analyte Sytem

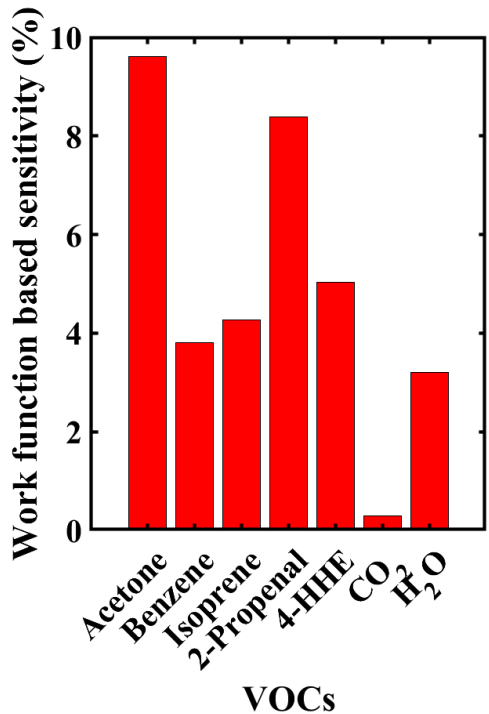

**Figure 8.** Work function-based sensitivity for different VOC adsorbed system

Work function ($\Phi$) for pristine and different VOC adsorbed structures was determined using the following formula [47]:

$$\Phi = V_\infty - E_F \qquad (4)$$

Where $V_\infty$ is the electrostatic potential, and $E_F$ is the fermi potential.

Sensors based on work function can achieve micro gas sensors with minimal power consumption, making them well-suited for cost-effective applications[48]. From **Table 1**, the work function of the pristine BeS monolayer is observed to be 6.56eV. Work function due to the interaction of ambient molecules such as $CO_2$ and $H_2O$ is 6.585 and 6.340 [34]. **Figure 8** illustrates the work function-based sensitivity, which was calculated using the following equation: -

$$S = \frac{|\Delta\Phi|}{\Phi} \times 100\% \qquad (5)$$

The increase in work function was more pronounced for the adsorbed structures of acetone and 2-propenal, measuring 9.603% and 8.38%, respectively. Conversely, the work function shifts induced by the adsorption of benzene, isoprene and 4HHE measured slightly higher at 3.81%, 4.27% and 5.03%, respectively, compared to the shift caused by the adsorption of ambient molecule $H_2O$. However, the impact of $CO_2$ is minimal due to its lower sensitivity.

### F. Conductivity of BeS-Analyte System

2D nanomaterial demonstrates promise as chemiresistive sensors due to their large surface area, charge transfer mechanisms and compatibility with flexible electronics to revolutionize gas sensing capabilities [49,50]. The Presence of particular VOCs can alter the conductive property of the pristine monolayer. The conductivity of a material is directly related to the bandgap by the following equation [28] :-

$$\sigma = A \, exp\left(\frac{E_g}{2k_BT}\right) \quad (6)$$

The following formula can be used to define the change in conductivity:

$$\%S\left(\frac{\delta\sigma}{\sigma}\right) = \frac{exp\left(-\frac{Eg(adsorbed)}{2k_BT}\right) - exp\left(-\frac{Eg(pristine)}{2k_BT}\right)}{exp\left(-\frac{Eg(pristine)}{2k_BT}\right)} \times 100 \quad (7)$$

Here, $k_B$ defines the Boltzmann constant and T is room temperature (298K). $E_g$(adsorbed) and $E_g$(pristine) are the bandgap of VOC adsorbed BeS monolayer and pristine BeS monolayer. **Table 1** shows that when 2-propenal is adsorbed, there's a significant 66.62% decrease in the bandgap. Similarly, the adsorption of acetone and 4HHE reduces the bandgap by 43.01% and 57.61%, respectively. These drastic changes in bandgap strongly affect conductivity. 21.97% reduction in bandgap also shows effect in conductivity change due to adsorption of isoprene. Also, it can be observed from **Table 2** that the percentage change in bandgap for these VOCs is significantly higher compared to that of other monolayers. However, when benzene is adsorbed, the bandgap only changes by 5.52%, which is still higher than the changes caused by interfering molecules like $H_2O$ and $CO_2$, which are only 0.04% and 1.63%, respectively.

### G. Recovery time

In sensing applications, recovery time refers to the duration it takes for the sensing materials to revert to its original state after interacting with target molecules. A shorter recovery time can enhance the efficiency and throughput of the biosensing system. Extended recovery time could result in residual molecules from prior measurements interfering with subsequent detections, potentially leading to inaccurate results. The amount of time required for recovery was determined by applying the Van't Hoff-Arrhenius equation.[51]

$$\tau = v_0^{-1} exp\left(\frac{-E_{ad}}{k_BT}\right) \quad (8)$$

where T is the temperature in kelvin, and $v_0$ is the attempt frequency. Prior studies established that the attempt frequency of benzene and acetone was $10^{13}$ Hz [52,53]. Based on previous research, it is reasonable to presume a similar attempt frequency for other VOCs. **Table 1** presents data regarding the recovery times of various molecules under visible light at a temperature of 298K. Due to physisorption, recovery time for benzene was lesser compared to other VOCs. It has been noted that ultraviolet (UV) radiation aids the sensor's recovery by lowering the desorption barrier [54]. Additionally, **Table 1** demonstrates that under UV light, the recovery time can be decreased at a temperature of 298K. During exposure to UV radiation, the attempt frequency was estimated to be approximately $10^{16}$ Hz [55]. From **Table 2**, it is evident that the recovery time for some VOCs on $SnS_2$ is less than nanoseconds. This short time is not suitable for VOC sensing. Conversely, metal-doped $MoS_2$, $SnS_2$, and $MoSe_2$ exhibit significantly longer recovery times for most scenarios, ranging from $3.26\times10^{12}$ to $6.56\times10^{31}$ seconds. In these instances, the material's reusability cannot be guaranteed. However, BeS demonstrates a moderate recovery time, ensuring the sensing material's reusability.

## H. Temperature Effect

Below a specific temperature threshold, the adsorption process is spontaneous and energetically favorable, aiding in the loading of VOC molecules onto the nanosheet through physisorption or chemisorption. Since DFT calculations are conducted at 0K, the desorption temperature for each VOC was determined for the real-world environmental assessment using the Van't Hoff equation [56].

$$T_D = \frac{E_{ad}}{k_B} \left[\frac{\Delta S}{R} - ln\frac{P}{P_0}\right]^{-1} \qquad (9)$$

In this equation, $E_{ad}$ represents the adsorption energy, and P stands for pressure, with a reference pressure $P_o$ set at 1 atm. R represents the universal gas constant and $\Delta S$ signifies the entropy change of the VOC. The variation of thermodynamic properties for all the VOCs is shown in **Figure S3**. **Table 1** indicates that at a pressure of 1 atm, the desorption temperatures for acetone, isoprene ,2-propenal and 4HHE are 411.16K, 315.83 K, 309.83K and 334.38K, respectively. Interestingly, the desorption temperatures are above the ambient temperature in these instances. However, for benzene, the desorption temperature is theoretically observed to be lower than room temperature, specifically 294.45 K.

## I. Effect of interfering molecules

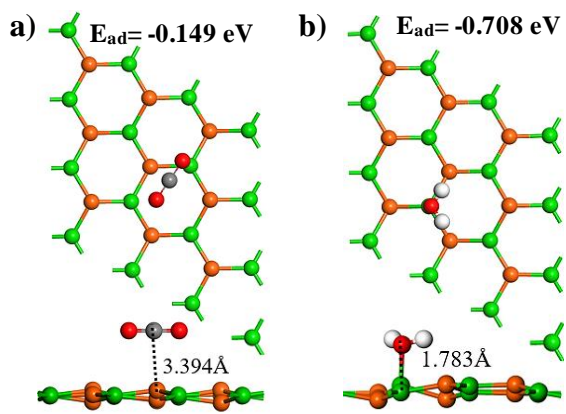

**Figure 9.** Optimized geometry (top and side view), (a) $CO_2$ and (b) $H_2O$ adsorbed structure

To ensure selectivity, it's essential to analyze the effects of ambient molecules such as carbon dioxide ($CO_2$) and humidity ($H_2O$) on the sensing mechanism. **Figures 9(a)** and **9(b)** depict the geometry of the adsorbed structure in the presence of these molecules. The adsorption of $CO_2$ has minimal impact on the pristine BeS monolayer, with a lower adsorption energy of -0.149 eV. Moreover, it doesn't alter the electronic band structure significantly, resulting in only a 1.63% change in the band gap. This is validated by the PDOS plot in **Figure 6(e)**, which shows no hybridization near the CBM or VBM and in the forbidden region. The charge transfer is notably low at 0.004|e|, indicating that $CO_2$ is physisorbed, with its effect being negligible. Similarly, the presence of humidity also slightly affects the electronic structure and PDOS, as depicted in **Figure 6(f)**. Notably, only a 0.04% change in bandgap is observed due to $H_2O$. However, the adsorption energy and charge transfer for water are higher compared to carbon dioxide, suggesting bond formation. Nevertheless, the adsorption energy of most VOCs remains higher than that of ambient molecules.

## J. Enhancing Sensing Response Through Strain

Previous studies have shown that applying strain has the potential to significantly enhance sensitivity by precisely manipulating the atomic scale of graphene nanosheets [57,58]. Due to the graphene-like hexagonal structure of the BeS monolayer, this study investigated its strain-induced sensing response. The percentage of strain can be calculated from the following formula

$$\epsilon = \frac{a - a_0}{a_0} \times 100 \qquad (10)$$

where $a_0$ is the lattice parameter of the BeS monolayer at no strain, and $a$ is the lattice parameter after introducing strain. Both uniaxial and biaxial strains have been applied to test the performance, as depicted in **Figure 10.** The adsorption structures of all VOCs exhibited a consistent trend in response to applied strain. Tensile strain led to a decrease in adsorption energy, prompting the desorption of molecules. Conversely, compressive strain notably enhanced the adsorption energy.

Uniaxial strain applied in both the a and b directions resulted in similar patterns of adsorption energy. For acetone and benzene, the uniaxial compressive strain demonstrated greater enhancement in adsorption energy compared to the biaxial strain. However, for 2-propenal, biaxial compressive strain showed better performance. Intriguingly, all types of compressive strains ranging from -2% to -4% resulted in enhanced adsorption. In conclusion, compressive strain within the range of -2% to -4% is optimal for enhancing the adsorption of VOCs on the BeS monolayer.

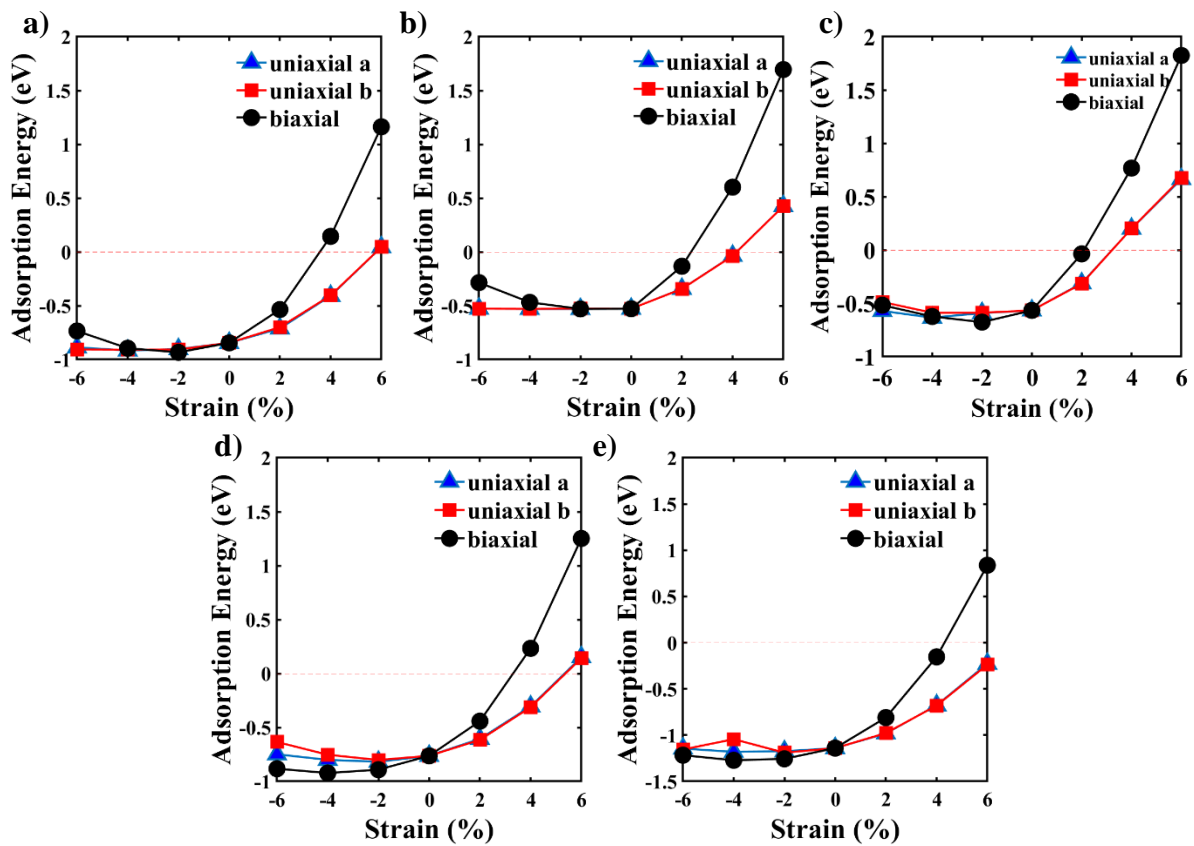

**Figure 10.** Adsorption energy versus applied strain for (a) acetone, (b) benzene, (c) isoprene, (d) 2-propenal and (e) 4-hydroxy hexanal adsorbed on BeS monolayer

**K. Enhancing Sensing Response Through Vertical Electric Field**

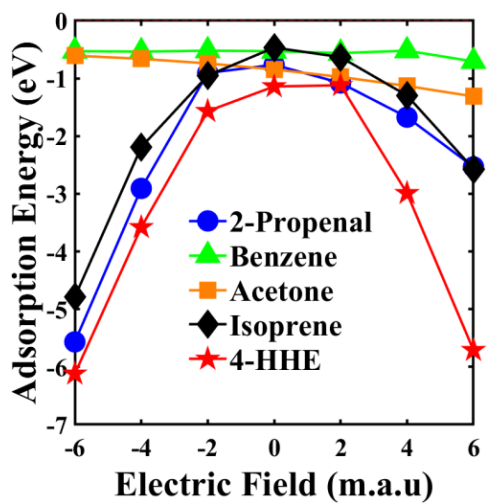

**Figure 11.** Adsorption energy versus applied electric field for acetone, benzene, isoprene, 2-propenal and 4HHE adsorbed BeS monolayer. The electric field is applied perpendicular to the BeS monolayer, with its positive direction oriented upward along the z-axis. (1 milli atomic unit (m.a.u) = $5.14 \times 10^6$ V/cm)

Prior studies showed Applying an electric field vertically significantly improves the sensing performance of the material [59,60]. **Figure 11** illustrates the impact of the vertically applied electric field. For acetone and benzene, a positive electric field strengthens the adsorption energy up to -1.858 eV and -1.028 eV, respectively, while a negative electric field reduces the adsorption temperature. In these cases, the electric field can be utilized to both adsorb and desorb volatile organic compound (VOC) molecules from the surface of the BeS monolayer. Interestingly, for 2-propenal, isoprene and 4HHE, both positive and negative electric fields increase the adsorption energy of the configuration. With a positive 6 m.a.u, the adsorption energy can reach -5.727 eV for 4HHE.

**L. Application Description**

Lung cancer patients display altered concentrations of various volatile compounds in their breath. The presence or absence of VOCs can indicate the likelihood of lung cancer [61]. The presence of the VOCs like hydro carbons and benzene derivatives, which are possible markers for lung cancer, is promising to detect the disease through breath analysis [62]. Among the VOCs, acetone serves as an initial biomarker for various diseases, including asthma, CKD, CLD, COPD, cystic fibrosis, diabetes, sleep apnea, malaria, and lung cancer. Similarly, isoprene acts as a biomarker for several diseases, including CLD, cystic fibrosis, diabetes, sleep apnea, and lung cancer [63]. Non-small cell lung cancer (NSCLC) and COPD patients, as well as control smokers, exhibited higher levels of isoprene in their exhaled breath [64]. Additionally, the presence of benzene has been observed in the breath of past smokers, indicating that cancer patients with a history of smoking have higher concentrations of benzene in their breath [65]. 2-propenal has strong correlation with lung cancer. Besides, diseases such as pulmonary fibrosis have shown some correlation with this VOC [66].

On the other hand 4HHE was exclusively found in lung cancer patients and was not detected in the air. Therefore, environmental air does not significantly affect the concentrations of this carbonyl VOC in exhaled breath samples [18,67]. Consequently, it can be inferred that these carbonyl compounds originate primarily from alveolar breath and that their concentrations increase might be able to selectively detect lung cancer.

Various types of VOCs may be present simultaneously, making the classification of specific disease via VOCs is quite challenging. Acetone and isoprene can act as the biomarker for most of lung related disease. Among the VOCs, 4HHE demonstrates a strong interaction with the BeS monolayer, exhibiting the highest adsorption energy compared to other VOCs. Consequently, the recovery time for 4HHE is longer than for other VOCs. Therefore, if the monolayer is exposed to UV light for a certain period, it might be possible that only 4HHE will remain on the monolayer, with all the characteristic indicating its presence. Additionally, applying a positive electric field bias may further enhance the interaction, as the energy of 4HHE becomes more favorable with the applied positive bias compared to other VOCs. The entire process is depicted in **Figure 12.** Since 4HHE is exclusively found in lung cancer effected patient's breath, the proposed monolayer may be able to distinguish lung cancer from other lung diseases. However, the BeS monolayer may not be as effective in specifically distinguishing diseases such as chronic liver disease, chronic obstructive pulmonary disease, cystic fibrosis, diabetes, sleep apnea, and malaria, as these conditions share common biomarkers like acetone and isoprene. This overlap makes it challenging for the BeS monolayer to accurately differentiate between them. Nonetheless, the BeS monolayer can serve as an initial screening tool, indicating the potential presence of one of these diseases.

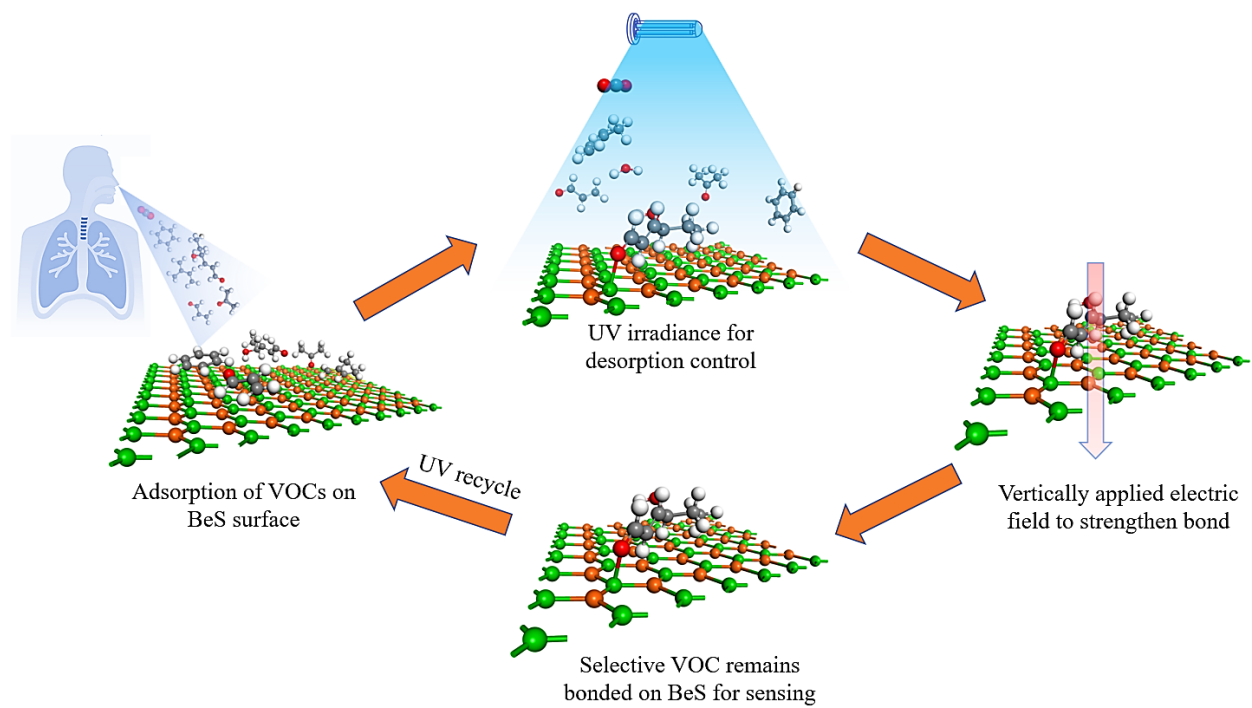

**Figure 12.** The proposed methodology of BeS monolayer for lung disease diagnosis.

**Table 2.** Recovery time, percentage of bandgap change with respect to pristine monolayer, and charge transfer for other monolayers.

| Monolayer | Recovery time(s) @298K | | | | %ΔE$_g$ | | | | |Q (e)| | | | |
|---|---|---|---|---|---|---|---|---|---|---|---|---|
| | C$_3$H$_6$O | C$_6$H$_6$ | C$_5$H$_8$ | C$_3$H$_4$O | C$_3$H$_6$O | C$_6$H$_6$ | C$_5$H$_8$ | C$_3$H$_4$O | C$_3$H$_6$O | C$_6$H$_6$ | C$_5$H$_8$ | C$_3$H$_4$O |
| **BeS** | 19.72 | 8.58×10$^{-5}$ | 2.08×10$^{-4}$ | 0.81 | 43.01 | 5.23 | 21.97 | 66.62 | 0.343 | 0.055 | 0.258 | 0.282 |
| **MoS$_2$[24]** | - | - | 9.16×10$^4$ | 27.24 | - | - | 0.617 | 1.234 | - | - | 0.02 | 0.08 |
| **SnS$_2$[26]** | 1.31×10$^{-12}$ | 3.76×10$^{-13}$ | 3.34×10$^{-13}$ | - | - | - | - | - | - | - | - | - |
| **WSe$_2$[27]** | 3.94×10$^{-7}$ | - | 166.27 | - | 6.67 | - | 3.33 | - | 0.004 | - | 1.031 | - |
| **C-MoS$_2$[24]** | - | - | 6.11×10$^{25}$ | 1.81×10$^{22}$ | - | - | 102.8 | 2.89 | - | - | 0.032 | 0.032 |
| **Al-MoSe$_2$[28]** | 1.82×10$^{18}$ | - | 6.47×10$^{21}$ | 3.26×10$^{12}$ | 23.53 | - | 25 | 14.7 | 0.230 | - | 0.060 | 0.160 |
| **Ni-MoS$_2$[29]** | - | 4.70×10$^{12}$ | 1.36×10$^{18}$ | 3.79×10$^{16}$ | - | 5.58 | 8.31 | 7.67 | - | 0.146 | 0.191 | 0.188 |
| **Sc-SnS$_2$[30]** | - | 7.77×10$^{36}$ | 6.56×10$^{31}$ | 4.91×10$^{11}$ | - | - | - | - | - | 0.173 | 0.249 | 0.014 |
| **Pd-SnS$_2$[26]** | 8.02×10$^{-4}$ | 2.66×10$^{-3}$ | 16.02 | - | 25.96 | 33.68 | 64.56 | - | - | - | - | - |
| **Ti$_3$C$_2$O$_2$[31]** | - | 8.14×10$^{-4}$ | 8.75×10$^{-3}$ | 4.58×10$^{-6}$ | - | - | - | - | - | 0.051 | 0.056 | 0.017 |
| **Ti$_3$C$_2$F$_2$[31]** | - | 1.47×10$^{-5}$ | 2.08×10$^{-4}$ | 5.18×10$^{-5}$ | - | - | - | - | - | 0.059 | 0.030 | 0.032 |
| **Ti$_3$C$_2$S$_2$[31]** | - | 9.28×10$^{-7}$ | 2.96×10$^{-4}$ | 5.10×10$^{-4}$ | - | - | - | - | - | 0.093 | 0.129 | 0.102 |
| **Ti$_3$C$_2$(OH)$_2$[31]** | - | 0.11 | 494.70 | 2.95×10$^{19}$ | - | - | - | - | - | 0.028 | 0.182 | 0.372 |
| **Ti$_3$C$_2$F(OH)$_2$[31]** | - | 3.3×10$^{-3}$ | 0.34 | 8.83×10$^3$ | - | - | - | - | - | 0.203 | 0.122 | 0.142 |

## Conclusion

In conclusion, this comprehensive study has elucidated the interaction of volatile organic compounds (VOCs) related to various lung diseases with BeS monolayer. Through detailed computational analyses, including electronic band structure analysis, phonon spectra calculations, evaluation of charge transfer, optical property, work function, recovery time assessments, and temperature and interference effects, valuable insights into the feasibility and efficiency of BeS for biosensing applications to detect VOC have been demonstrated. The findings shows significant changes in bandgap upon VOC adsorption, with reductions of 66.62%, 43.01%, and 57.61% observed for 2-propenal, acetone, and 4HHE, respectively. Sensing based on work function was found to be over 5% for acetone, 2-propenal and 4HHE. Regarding optical properties, benzene and isoprene adsorbed structures displayed peak absorbance around 120nm, enabling their selective detection, while within the visible region (400nm to 700nm), all volatile organic compounds exhibited over 20% reflectivity, distinguishing them from interfering molecules. Moreover, recovery times ranging from microseconds to several seconds were evaluated, highlighting BeS's moderate recovery time compared to other 2D materials. Additionally, applying compressive strain in the range of -2 to -4 % and positive vertical electric fields demonstrated promising enhancements in adsorption energy thus improving interaction. The classification of specific VOCs is challenging when multiple VOCs are present simultaneously. Among these, 4HHE exhibits the strongest interaction with the BeS monolayer, having the highest adsorption energy and longest recovery time. Applying a positive electric field bias can further enhance this interaction, increasing the probability of selective detection. These characteristics underscore BeS's potential as a sensing materials for detection of breath biomarkers responsible for particular lung disease like lung cancer.

**Data Availability**

The data supporting this article have been included as part of the ESI.

**Authors Contribution**

Sudipta Saha contributed to the conceptualization, methodology, visualization, software, investigation, drafting of the original manuscript and editing the manuscript. Md. Kawsar Alam was involved in conceptualization, methodology, visualization, resource management, original draft writing, reviewing and editing the manuscript, and provided supervision throughout the research process.

**Conflict of Interest**
There are no conflicts to declare.